\documentclass[aps,prl,twocolumn,groupedaddress]{revtex4}
\def\lsim{~\rlap{\raise .2ex\hbox{$<$}}{\lower .9ex\hbox{$\sim$}}}
\def\gsim{~\rlap{\raise .2ex\hbox{$>$}}{\lower .9ex\hbox{$\sim$}}}
\usepackage{graphicx}
\usepackage{bm}

\begin{document}
\title{Electron Temperature Evolution in Expanding Ultracold Neutral Plasmas}

\author{P. Gupta, S. Laha, C. E. Simien, H. Gao, J. Castro, and T. C. Killian}
\affiliation{Rice University, Department of Physics and Astronomy,
Houston, Texas, 77005}
\author{T. Pohl}
\affiliation{ITAMP, Harvard-Smithsonian Center for Astrophysics, 60
Garden Street, Cambridge, MA 02138}
\date{\today}

\begin{abstract}
We have used the free expansion of ultracold neutral plasmas  as a
time-resolved probe of electron temperature. A combination of
experimental measurements of the ion expansion velocity and
numerical simulations characterize the crossover from an
elastic-collision regime at low initial $\Gamma_e$, which is
dominated by adiabatic cooling of the electrons, to the regime of
high $\Gamma_e$ in which inelastic processes drastically heat the
electrons. We identify the time scales and relative contributions of
various processes, and experimentally show the importance of
radiative decay and disorder-induced electron heating for the first
time in ultracold neutral plasmas.

\end{abstract}


\maketitle

 Ultracold neutral plasmas (UNPs) \cite{kil07} occupy
an exotic regime of plasma physics in which electron and ion
temperatures are orders of magnitude colder than in conventional
neutral plasmas. The electron temperature in these systems evolves
under the influence of many factors, which can occur on very
different time scales, such as disorder-induced heating
\cite{kon02}, three-body recombination \cite{mke69}, and adiabatic
cooling \cite{rha03,ppr04PRA}. The relative importance of the
various effects depends critically upon initial conditions, and this
has complicated the experimental study of the electron temperature
\cite{kkb00,rfl04,cdd05,cdd05physplasmas,fzr06} and lead to much
theoretical debate  \cite{tya00,rha03,mck02,kon02,ppr04PRA}. We
present here detailed experimental measurements and numerical
simulations that untangle the time scales and contributions of the
various competing effects and characterize the transition from
elastic-collision-dominated to inelastic-collision-dominated
behavior.

 UNPs are  of fundamental interest because they can be in
 or near the strongly
coupled regime, which is characterized by the existence of spatial
correlations between particles and a Coulomb coupling parameter
$\Gamma = e^2/(4\pi\varepsilon_0 a k_\mathrm{B} T )> 1$, where $T$
refers to the temperature of the particles and $a=(4\pi n/3)^{-1/3}$
is the Wigner-Seitz radius.  Ions in UNPs equilibrate with $\Gamma_i
\sim 3$ \cite{scg04,cdd05}. The initial electron temperature is
under experimental control and can be set such that a naive
calculation of $\Gamma_e$ suggests that electrons are also strongly
coupled. However, electrons rapidly leave the strongly coupled
regime due to various heating mechanisms \cite{rha03,kon02,mck02}
that are central to studies presented here.

UNPs are created by photoionizing laser-cooled Sr atoms \cite{nsl03}
just above the ionization threshold  with a 10\,ns laser pulse. The
ion temperature is initially a few millikelvin, which is similar to
the temperature of the laser-cooled neutral atoms, but ions heat
within one microsecond to about 1\,K due to disorder-induced heating
\cite{mur01,scg04}. The initial electron kinetic energy ($E_e$)
equals the difference between the energy of the ionizing photon and
the ionization threshold. With a tunable pulsed-dye laser,
$2E_e/3k_\mathrm{B}$ can be set from 1-1000\,K. Electrons thermalize
locally  within 100\, ns and globally within $1\,\mu$s \cite{rha03}.
Simple equilibration would set the initial $T_e=2E_e/3k_\mathrm{B}$,
but we will discuss processes that can change $T_e$.

The plasma density follows the profile of the neutral atom cloud. By
adjusting the laser-cooling parameters and imaging the cloud in two
perpendicular axes, we ensure that the plasma has a spherically
symmetric Gaussian profile, $n(r)=n_0e^{-r^2/2\sigma^2}$. Deviations
from spherical symmetry, \textit{e.g.}
$(\sigma_x-\sigma_y)/\sigma_x$, are less than 5\%. Typically, the
initial $n_0$ is  $\sim$$10^{16}$\,m$^{-3}$ and $\sigma \sim 1$\,mm.
The plasma is quasineutral ($n_i \sim n_e$)  with the Debye length
$\lambda_D=(\varepsilon_0 k_\mathrm{B} T_e/ n_{e} e^2)^{1/2} \ll
\sigma$, where $n_i$ and $n_e$ refer to ion and electron density
respectively. UNPs are unconfined and expand into the surrounding
vacuum, and quasineutrality is maintained during the expansion.

Electron temperature evolution during expansion of ultracold plasmas
has been studied using various techniques. Electron plasma
oscillations \cite{kkb00} measured the average density in order to
obtain the rms radial terminal velocity of the ions. This showed
that essentially all the initial electron energy is converted to ion
expansion energy. In addition,
plasmas with lower $E_e$ and higher $n_0$ (which would imply
$\Gamma_e\gsim 1$ \cite{electrontempgammenote}) resulted in an
anomalously fast expansion. Numerical simulations \cite{rha03}
showed that various electron heating mechanisms explained the
result. However, the relationship between density and the electron
plasma oscillation used in \cite{kkb00} has been called into
question in subsequent work \cite{bsp03,fzr06}. Ref.\ \cite{rfl04}
probed $T_e$ by measuring the fraction of electrons escaping the
plasma during a small electric field pulse and inferred that for
$10\,\mathrm{K} <2E_e/3k_\mathrm{B}<300$\,K and $n_0\sim 5\times
10^{14}$\,m$^{-3}$,
 electrons approach a narrow range
 of electron temperatures ($20\,\mathrm{K}<T_e<40\,\mathrm{K}$) about 5 $\mu$s after
 photoionization.
  Spatially resolved  fluorescence
detection of the ions \cite{cdd05,cdd05physplasmas}  in a
cylindrical plasma measured expansion energies  similar to what was
observed in \cite{kkb00}, but they found significant deviations from
theoretical predictions that perhaps arose because of the lack of
spherical symmetry.
Tonks-Dattner modes, which have resonant frequencies that are
sensitive to $T_e$,  were recently observed  in  UNPs \cite{fzr06}.
They found agreement between data and a model that assumed constant
$T_e$ over $40\,\mu$s of evolution. So the  electron temperature
evolution in UNPs is obviously a complicated problem that remains
unsettled.

 The results presented here have
several advantages over previous studies. We have a spherically
symmetric plasma that allows the application of exact analytic
results. Doppler broadening of the ion optical absorption spectrum
\cite{scg04} provides a calibrated, model-independent measure of the
ion velocity and overall plasma expansion  with excellent temporal
resolution stretching from the phase of initial ion acceleration to
the onset of terminal velocity. A combination of experiment and
numerical simulation allows the contributions of various electron
heating and cooling mechanisms to be separated as never before, and
we find excellent agreement between experiment and theory with no
adjustable parameters. We also present a systematic study of the
full spectrum of dynamics observed in current UNP experiments.

For an experimental probe, the Doppler width of the ion absorption
spectrum \cite{scg04} for the entire plasma  measures
$\sqrt{\langle(\textbf{v}\cdot \textbf{\^{z}})^2\rangle} \equiv
\mathrm{\textsl{v}}_{i,rms}$ \cite{electrontempvirmsnote}, where
$\textbf{\^{z}}$ is the laser propagation direction, $\mathbf{v}$ is
the total ion velocity including random thermal motion and expansion
($\mathbf{u}$), the angled brackets refer to an average over the
plasma density and velocity distribution. In a quasineutral Gaussian
plasma such as a UNP, the electron temperature
 can be found from   measurements of $\mathrm{\textsl{v}}_{i,rms}$
 due to its sensitivity to $\mathbf{u}$ and
 the fact that the expansion
 acceleration,
\begin{equation} \label{eq1}
\dot{\bf u}=-\frac{k_{\rm B}\left(T_{e}+T_{
i}\right)}{m_{i}}\frac{\nabla n}{n},
\end{equation}
arises from thermal pressure \cite{dse98,kkb00,rha03,ppr04PRA}.


For high $E_e$ and low $n_0$, which we denote as the ``elastic
collisional regime" ($\Gamma_e<0.1$ \cite{electrontempgammenote}),
all collisional processes in a UNP are elastic, the plasma expands
adiabatically, and electrons cool \cite{ppr04}. This leads to a
self-similar expansion that preserves the Gaussian phase-space
distributions and is described
 by an analytic solution of the Vlasov equations \cite{rha03,ppr04PRA} that was
originally derived for short-pulse laser experiments
\cite{bku98,dse98,kby03} and applied to UNPs in
\cite{rha03,ppr04PRA}.
UNPs provide the first clean realization of this analytic solution,
and this was shown experimentally in \cite{lgg07}.
The Vlasov equations do not include a collision term, which is
appropriate because such a term vanishes for a Maxwell-Boltzmann
velocity distribution. So in this sense, the expansion in the
elastic collisional regime can also be called ``collisionless"
\cite{dse98}.

 In this
regime, $\mathrm{\textsl{v}}_{i,rms}$ is given by \cite{lgg07}
\begin{eqnarray} \label{eq3}
\mathrm{\textsl{v}}_{i,rms} = \sqrt{\frac{k_{\rm B}}{m_{
i}\tau_{exp}^2}\left[t^2\left(T_{ e}+T_{ i}\right)+\tau_{
exp}^2T_i\right]}.
\end{eqnarray}
The characteristic expansion time $\tau_{ exp}$ is given by $\tau_{
exp}=\sqrt{m_{ i}\sigma(0)^2/k_{\rm B} [T_{ e}(0)+T_{ i}(0)]}$, and
the electron and ion temperatures follow
\begin{equation}\label{adibatictemperatures}
T_{ e/i}=T_{ e/i}(0)/(1+t^2/\tau_{ exp}^2).
\end{equation}

\begin{figure}[t]
\centering
\includegraphics[clip=true,angle=-90,width=3.25in, totalheight=1.450in, trim=20 00 290 0]{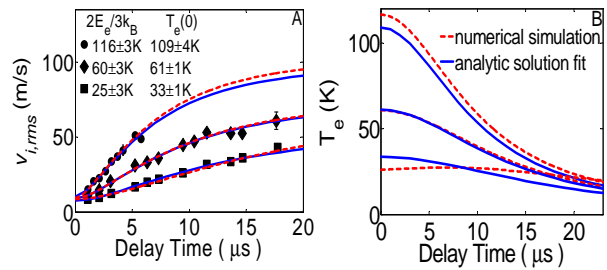}
\caption[High $2E_e/3k_\mathrm{B}$]{\label{highTedata} Expansion
velocity (A) and electron temperature (B) for low $\Gamma_e$ showing
little or no electron heating effects. (A) The initial peak
densities and sizes are $n_0 = 3.5\times 10^{15}\, \mathrm{m}^{-3}$
and $\sigma(0) = 1$\, mm. The self-similar analytic solution (dark
solid line) provides an excellent fit of the data, with fit $T_e(0)$
indicated in the legend. The full numerical simulation (dashed line)
describes the data with no adjustable parameters.   (B) Electron
temperature evolutions (Eq.\ \ref{adibatictemperatures}) are
determined from parameters of the fits to
$\mathrm{\textsl{v}}_{i,rms}$ or by numerical simulation. For
$2E_e/3k_\mathrm{B}=25$\,K, the analytic solution fit and numerical
simulation for $T_e(t)$ deviate significantly, showing the
importance of inelastic collision effects.}
\end{figure}

Figure\ \ref{highTedata}A shows the ion velocity evolution for UNPs
in the elastic collisional regime. Fits using Eq.\ \ref{eq3} take
$T_{i/e}(0)$ as fit parameters, while $\sigma(0)$ is fixed to the
value found from the images. For higher $E_e$ in Fig.\
\ref{highTedata}, the fit is excellent and the extracted values of
$T_e(0)$ yield $2E_e/3k_\mathrm{B}$ within experimental uncertainty,
confirming that inelastic collisions are negligible in this regime.
The underlying electron temperature evolution is shown in Fig.\
\ref{highTedata}B. As expected, $T_e$ drops due to adiabatic cooling
because there is no significant electron heating.

For  $2E_e/3k_\mathrm{B}=25$\,K in Fig.\ \ref{highTedata}, the fit
$T_e(0)=33\,$K exceeds the expected value. This provides evidence
that inelastic processes are modifying the electron temperature.
Many processes are expected to contribute.
Within tens of nanoseconds,
disorder-induced heating (DIH) \cite{kon02,electrontemptlnote}
increases electron kinetic energy  by
as much as several kelvin above $E_e$ in the range of our initial
conditions. DIH is the conversion of potential energy into kinetic
as spatial correlations develop, which is an effect of strong
coupling that has been observed for ions in UNPs \cite{scg04}.
Three-body recombination (TBR) \cite{mke69} populates Rydberg levels
bound by $\sim k_\mathrm{B} T_e$ and heats the free electrons. The
total TBR rate varies  as $T_e^{-9/2}$ and can be very rapid in
ultracold systems \cite{klk01}.
Rydberg-electron collisions (REC) \cite{mke69, rha03}  can transfer
Rydberg atoms to more deeply bound levels and heat electrons
further, while radiative decay (RD) of the Rydberg atoms mitigates
this heating effect. Direct cooling of electrons through
equilibration with ions is negligibly slow for these experiments
because of the large ion-electron mass difference \cite{ppr05JPB}.

To understand the interplay between these various effects, we
performed numerical simulations, taking into account all relevant
heating and cooling mechanisms. Our description is based on a
particle-in-cell simulation of the ions and treats the electrons
adiabatically as a fluid in a constantly changing
equilibrium state \cite{rha03,ppr04PRA}. DIH 
due to particle correlations is accounted for in the initial
conditions of both plasma components \cite{mur01,kon02}, assuming a
homogeneous $T_{\rm e}$ for the electrons and a homogeneous
$\Gamma_i$ for the ions. This treatment neglects the influence of
correlations on later stages of  the plasma dynamics, which may be a
concern because UNP ions are strongly coupled \cite{scg04,cdd05},
but \cite{ppr04PRA} showed that the effects of ion-ion correlations
on the expansion are negligible. Finally we use a Monte-Carlo
treatment to describe the formation of Rydberg atoms and their
subsequent binding energy evolution, employing known expressions for
the rates of TBR and REC \cite{mke69} and RD \cite{bsa77}. The TBR
rates \cite{mke69} are well-confirmed experimentally at high
temperatures, but their low temperature validity has been questioned
\cite{hah97,hah00,mty98} as it ultimately has to break down for $T_{
e}\rightarrow0$ due to its strong $\propto T_{\rm e}^{-9/2}$
temperature divergence. By comparing experiments and calculations
the present study
 indirectly tests  TBR theory over a wide range of
temperatures.

The numerical simulation reproduces the expansion dynamics in Fig.\
\ref{highTedata} with no adjustable parameters, and it shows  that
$2E_e/3k_\mathrm{B}=25$\,K is on the border of the ``inelastic
collisional regime", which is defined here as  initial $\Gamma_e>
0.1$ \cite{electrontempgammenote}. For this data set, $T_e(t)$ stays
roughly constant over the observed evolution time and is not
well-described by the analytic solution (Eq.\
\ref{adibatictemperatures}). Only TBR and REC processes must be
included in the simulation to accurately describe the expansion.


\begin{figure}[ht]
\centering
\includegraphics[clip=true,angle=-90,width=3.5in,totalheight=1.45in,trim=20 00 290 0]{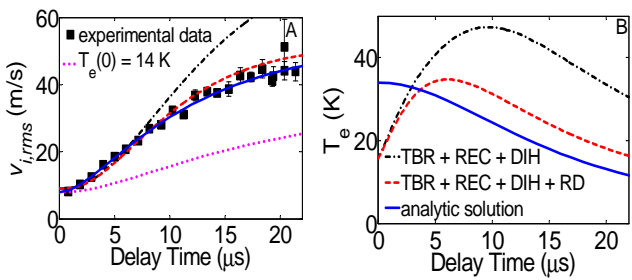}
\caption[Low $2E_e/3k_\mathrm{B}$]{(A) Expansion velocity  and (B)
electron temperature (B) for  moderate initial $\Gamma_e=0.2$,
($2E_e/3k_B=14$\,K, $\sigma(0)=0.9$\,mm, and $n_0(0)=7\times
10^{15}$\,m$^{-3}$). (A) A fit of the expansion to the analytic
solution (Eq.\ \ref{eq3}, solid line) yields $T_e(0)=34\,$K, which
reflects much faster expansion than expected for elastic-collisional
dynamics with $2E_e/3k_B=14$\,K (dotted line).  Only including TBR
and REC in the simulation (dot-dash line)  overestimates the
heating. Including RD as well (dashed line) brings theory into good
agreement with experiment. Including DIH increases the initial
electron temperature by about $3\,$K, but has little effect on the
velocity fit. (B) The fit $T_e(0)$ approximately equals the maximum
of the actual $T_e(t)$ obtained with simulation. }\label{lowTedata}
\end{figure}

Fig.~\ref{lowTedata}A displays characteristic plasma dynamics for a
UNP with lower initial electron energy and higher $\Gamma_e$ that is
in the inelastic collisional regime. The expansion is much faster
than expected for elastic-collisional-regime dynamics with
$T_e(0)=2E_e/3k_B=14$\,K.
Incorporating only TBR and REC in the simulation considerably
overshoots the observed ion velocity evolution because it
overestimates electron heating. Inclusion of RD, which transfers
Rydberg atoms to more deeply bound states without heating the plasma
electrons, produces excellent agreement with experiment without
adjustable parameters.

The analytic  solution (Eq.\ \ref{eq3}) provides a surprisingly
accurate description of observed ion velocities, but the simulation
shows (Fig.~\ref{lowTedata}B) that the extracted $T_e(0)$ is only a
phenomenological parameter. Physically, TBR and REC heat the
electrons in the first $\sim 5$\,$\mu$s, although it is mitigated by
RD. The increasing $T_e$, and to a lesser extent decreasing density,
slows TBR, so that adiabatic cooling dominates at later times. The
fit $T_e(0)$ gives a rough estimate of the maximum electron
temperature, $T_{e,max}$.

\begin{figure}[hb]
\centering
\includegraphics[clip=true,angle=-90,width=3.5in,totalheight=1.45in,trim=20 00 290 0]{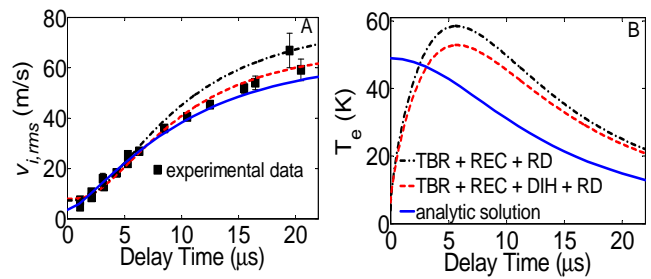}
\caption[ ] {(A) Ion velocity and (B) electron temperature for
$\Gamma_e \sim 1$, ($2E_e/3k_B=4$\,K, $n_0(0)=8\times
10^{15}\,\mathrm{cm}^{-3}$, $\sigma(0)=1.0\,\mathrm{mm}$). (A) The
numerical simulation including all heating effects matches the data
with no adjustable parameters. For this relatively high density, the
initial heating from DIH slows TBR and REC enough to produce an
observable effect. The analytic solution fit is poor, but yields
$T_e(0)=49\pm 2$\,K. (B) The electron temperature shows drastic
heating at early times and $T_e(0)$ no longer provides a good
estimate of $T_{e,max}$. {\label{lowTe}}}
\end{figure}

Further decrease of $E_e$ or increase in density pushes further into
the inelastic collisional regime where a naive calculation using
$T_e=2E_e/3k_\mathrm{B}$ implies $\Gamma_{ e}\gsim 1$.
In this regime, fits using the analytic expansion expression fail to
reproduce the data (Fig.\ \ref{lowTe}).  Simulations show that the
approximation of a self-similar Gaussian expansion also becomes poor
due to the large fraction of ions that undergo TBR and the higher
rate for this process in the higher density central region of the
plasma.
The  measured ion expansion velocity indicates the occurrence of
extreme electron heating  from $2E_e/3k_B=4$\,K. For this relatively
high-density sample DIH makes a significant contribution. It quickly
raises $T_e$, which slows recombination and leads to a lower
$T_{e,max}=53$\,K. The agreement between data and simulation
indicates no significant deviation from classical TBR theory
\cite{mke69}.

\begin{figure}[ht]
\centering
\includegraphics[clip=true,angle=-90,width=3.0in, trim=0 0 0 0]{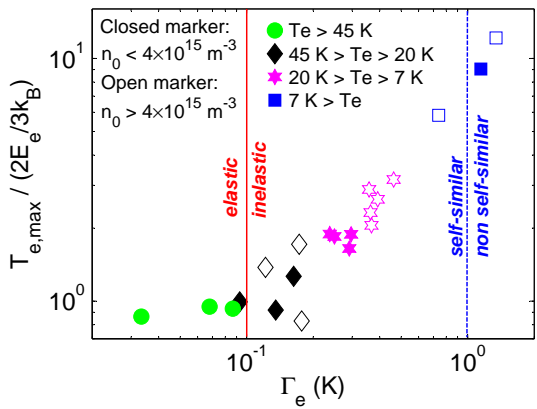}
\caption{Electron heating summary. The plasma dynamics are
parameterized well by $\Gamma_e(n_0,E_e)$. $T_{e,max}$ is
 the maximum electron temperature attained during
the evolution. The heating is negligible in the elastic collisional
regime ($\Gamma_e< 0.1$ \cite{electrontempgammenote}), and
$T_{e,max}/(2E_e/3k_B)\approx 1$. In the  inelastic collisional
regime, the heating becomes significant and increases with
increasing density and decreasing $2E_e/3k_\mathrm{B}$. Beyond
$\Gamma_e \approx 1$, the expansion ceases to be self-similar.
}\label{gammaefig}
\end{figure}

 Figure \ref{gammaefig} summarizes our results for
 electron heating in UNPs.
 The data
 is organized
 according to initial $\Gamma_e$ \cite{electrontempgammenote}, and
 it
displays a clear trend in the heating of the electrons as
previously
 observed in \cite{kkb00,cdd05physplasmas}.
The onset of heating occurs at the crossover between the inelastic
and elastic collisional regime. For initial $\Gamma_e \gsim 1$, the
expansion ceases to be self-similar.


We have described the use of spectroscopic measurements of the ion
velocity and numerical simulations as a time-resolved probe of the
electron dynamics of UNPs.  It resolves outstanding questions
regarding the evolution of the electron temperature, and shows that
the dynamics vary greatly for different  initial electron kinetic
energy and plasma density. This work identifies the relative
contributions and timescales of various processes, demonstrates the
importance of radiative decay and disorder-induced electron heating
for the first time in UNPs, and shows no discrepancy between
observed and theoretical rates of TBR.

This work was supported by the National Science Foundation (Grant
PHY-0355069 and a grant for the Institute of Theoretical Atomic,
Molecular and Optical Physics (ITAMP) at Harvard University and
Smithsonian Astrophysical Observatory) and the David and Lucille
Packard Foundation.

\begin{thebibliography}{31}
\expandafter\ifx\csname natexlab\endcsname\relax\def\natexlab#1{#1}\fi
\expandafter\ifx\csname bibnamefont\endcsname\relax
  \def\bibnamefont#1{#1}\fi
\expandafter\ifx\csname bibfnamefont\endcsname\relax
  \def\bibfnamefont#1{#1}\fi
\expandafter\ifx\csname citenamefont\endcsname\relax
  \def\citenamefont#1{#1}\fi
\expandafter\ifx\csname url\endcsname\relax
  \def\url#1{\texttt{#1}}\fi
\expandafter\ifx\csname urlprefix\endcsname\relax\def\urlprefix{URL }\fi
\providecommand{\bibinfo}[2]{#2}
\providecommand{\eprint}[2][]{\url{#2}}

\bibitem[{\citenamefont{Killian}(2007)}]{kil07}
\bibinfo{author}{\bibfnamefont{T.~C.} \bibnamefont{Killian}},
  \bibinfo{journal}{{Science}} \textbf{\bibinfo{volume}{316}},
  \bibinfo{pages}{705} (\bibinfo{year}{2007}).

\bibitem[{\citenamefont{Kuzmin and O'Neil}(2002)}]{kon02}
\bibinfo{author}{\bibfnamefont{S.~G.} \bibnamefont{Kuzmin}} \bibnamefont{and}
  \bibinfo{author}{\bibfnamefont{T.~M.} \bibnamefont{O'Neil}},
  \bibinfo{journal}{{ Phys. Plasmas}} \textbf{\bibinfo{volume}{9}},
  \bibinfo{pages}{3743} (\bibinfo{year}{2002}).

\bibitem[{\citenamefont{Mansbach and Keck}(1969)}]{mke69}
\bibinfo{author}{\bibfnamefont{P.}~\bibnamefont{Mansbach}} \bibnamefont{and}
  \bibinfo{author}{\bibfnamefont{J.}~\bibnamefont{Keck}}, \bibinfo{journal}{{
  Phys. Rev.}} \textbf{\bibinfo{volume}{181}}, \bibinfo{pages}{275}
  (\bibinfo{year}{1969}).

\bibitem[{\citenamefont{Pohl et~al.}(2004{\natexlab{a}})\citenamefont{Pohl,
  Pattard, and Rost}}]{ppr04PRA}
\bibinfo{author}{\bibfnamefont{T.}~\bibnamefont{Pohl}},
  \bibinfo{author}{\bibfnamefont{T.}~\bibnamefont{Pattard}}, \bibnamefont{and}
  \bibinfo{author}{\bibfnamefont{J.~M.} \bibnamefont{Rost}},
  \bibinfo{journal}{{ Phys. Rev. A}} \textbf{\bibinfo{volume}{70}},
  \bibinfo{pages}{033416} (\bibinfo{year}{2004}{\natexlab{a}}).

\bibitem[{\citenamefont{Robicheaux and Hanson}(2003)}]{rha03}
\bibinfo{author}{\bibfnamefont{F.}~\bibnamefont{Robicheaux}} \bibnamefont{and}
  \bibinfo{author}{\bibfnamefont{J.~D.} \bibnamefont{Hanson}},
  \bibinfo{journal}{{ Phys. Plasmas}} \textbf{\bibinfo{volume}{10}},
  \bibinfo{pages}{2217} (\bibinfo{year}{2003}).

\bibitem[{\citenamefont{Kulin et~al.}(2000)\citenamefont{Kulin, Killian,
  Bergeson, and Rolston}}]{kkb00}
\bibinfo{author}{\bibfnamefont{S.}~\bibnamefont{Kulin}},
  \bibinfo{author}{\bibfnamefont{T.~C.} \bibnamefont{Killian}},
  \bibinfo{author}{\bibfnamefont{S.~D.} \bibnamefont{Bergeson}},
  \bibnamefont{and} \bibinfo{author}{\bibfnamefont{S.~L.}
  \bibnamefont{Rolston}}, \bibinfo{journal}{{ Phys. Rev. Lett.}}
  \textbf{\bibinfo{volume}{85}}, \bibinfo{pages}{318} (\bibinfo{year}{2000}).

\bibitem[{\citenamefont{Roberts et~al.}(2004)\citenamefont{Roberts, Fertig,
  Lim, and Rolston}}]{rfl04}
\bibinfo{author}{\bibfnamefont{J.~L.} \bibnamefont{Roberts}},
  \bibinfo{author}{\bibfnamefont{C.~D.} \bibnamefont{Fertig}},
  \bibinfo{author}{\bibfnamefont{M.~J.} \bibnamefont{Lim}}, \bibnamefont{and}
  \bibinfo{author}{\bibfnamefont{S.~L.} \bibnamefont{Rolston}},
  \bibinfo{journal}{{Phys. Rev. Lett.}} \textbf{\bibinfo{volume}{92}},
  \bibinfo{pages}{253003} (\bibinfo{year}{2004}).

\bibitem[{\citenamefont{Cummings
  et~al.}(2005{\natexlab{a}})\citenamefont{Cummings, Daily, Durfee, and
  Bergeson}}]{cdd05}
\bibinfo{author}{\bibfnamefont{E.~A.} \bibnamefont{Cummings}},
  \bibinfo{author}{\bibfnamefont{J.~E.} \bibnamefont{Daily}},
  \bibinfo{author}{\bibfnamefont{D.~S.} \bibnamefont{Durfee}},
  \bibnamefont{and} \bibinfo{author}{\bibfnamefont{S.~D.}
  \bibnamefont{Bergeson}}, \bibinfo{journal}{{ Phys. Rev. Lett.}}
  \textbf{\bibinfo{volume}{95}}, \bibinfo{pages}{235001}
  (\bibinfo{year}{2005}{\natexlab{a}}).

\bibitem[{\citenamefont{Cummings
  et~al.}(2005{\natexlab{b}})\citenamefont{Cummings, Daily, Durfee, and
  Bergeson}}]{cdd05physplasmas}
\bibinfo{author}{\bibfnamefont{E.~A.} \bibnamefont{Cummings}},
  \bibinfo{author}{\bibfnamefont{J.~E.} \bibnamefont{Daily}},
  \bibinfo{author}{\bibfnamefont{D.~S.} \bibnamefont{Durfee}},
  \bibnamefont{and} \bibinfo{author}{\bibfnamefont{S.~D.}
  \bibnamefont{Bergeson}}, \bibinfo{journal}{{ Phys. Plasmas}}
  \textbf{\bibinfo{volume}{12}}, \bibinfo{pages}{123501}
  (\bibinfo{year}{2005}{\natexlab{b}}).

\bibitem[{\citenamefont{Fletcher et~al.}(2006)\citenamefont{Fletcher, Zhang,
  and Rolston}}]{fzr06}
\bibinfo{author}{\bibfnamefont{R.~S.} \bibnamefont{Fletcher}},
  \bibinfo{author}{\bibfnamefont{X.~L.} \bibnamefont{Zhang}}, \bibnamefont{and}
  \bibinfo{author}{\bibfnamefont{S.~L.} \bibnamefont{Rolston}},
  \bibinfo{journal}{{ Phys. Rev. Lett.}} \textbf{\bibinfo{volume}{96}},
  \bibinfo{pages}{105003} (\bibinfo{year}{2006}).

\bibitem[{\citenamefont{Tkachev and Yakovlenko}(2000)}]{tya00}
\bibinfo{author}{\bibfnamefont{A.~N.} \bibnamefont{Tkachev}} \bibnamefont{and}
  \bibinfo{author}{\bibfnamefont{S.~I.} \bibnamefont{Yakovlenko}},
  \bibinfo{journal}{{ Quantum Electronics}} \textbf{\bibinfo{volume}{30}},
  \bibinfo{pages}{1077} (\bibinfo{year}{2000}).

\bibitem[{\citenamefont{Mazevet et~al.}(2002)\citenamefont{Mazevet, Collins,
  and Kress}}]{mck02}
\bibinfo{author}{\bibfnamefont{S.}~\bibnamefont{Mazevet}},
  \bibinfo{author}{\bibfnamefont{L.~A.} \bibnamefont{Collins}},
  \bibnamefont{and} \bibinfo{author}{\bibfnamefont{J.~D.} \bibnamefont{Kress}},
  \bibinfo{journal}{{ Phys. Rev. Lett.}} \textbf{\bibinfo{volume}{88}},
  \bibinfo{pages}{55001} (\bibinfo{year}{2002}).

\bibitem[{\citenamefont{Simien et~al.}(2004)\citenamefont{Simien, Chen, Gupta,
  Laha, Martinez, Mickelson, Nagel, and Killian}}]{scg04}
\bibinfo{author}{\bibfnamefont{C.~E.} \bibnamefont{Simien}},
  \bibinfo{author}{\bibfnamefont{Y.~C.} \bibnamefont{Chen}},
  \bibinfo{author}{\bibfnamefont{P.}~\bibnamefont{Gupta}},
  \bibinfo{author}{\bibfnamefont{S.}~\bibnamefont{Laha}},
  \bibinfo{author}{\bibfnamefont{Y.~N.} \bibnamefont{Martinez}},
  \bibinfo{author}{\bibfnamefont{P.~G.} \bibnamefont{Mickelson}},
  \bibinfo{author}{\bibfnamefont{S.~B.} \bibnamefont{Nagel}}, \bibnamefont{and}
  \bibinfo{author}{\bibfnamefont{T.~C.} \bibnamefont{Killian}},
  \bibinfo{journal}{{Phys. Rev. Lett.}} \textbf{\bibinfo{volume}{92}},
  \bibinfo{pages}{143001} (\bibinfo{year}{2004}).

\bibitem[{\citenamefont{Nagel et~al.}(2003)\citenamefont{Nagel, Simien, Laha,
  Gupta, Ashoka, and Killian}}]{nsl03}
\bibinfo{author}{\bibfnamefont{S.~B.} \bibnamefont{Nagel}},
  \bibinfo{author}{\bibfnamefont{C.~E.} \bibnamefont{Simien}},
  \bibinfo{author}{\bibfnamefont{S.}~\bibnamefont{Laha}},
  \bibinfo{author}{\bibfnamefont{P.}~\bibnamefont{Gupta}},
  \bibinfo{author}{\bibfnamefont{V.~S.} \bibnamefont{Ashoka}},
  \bibnamefont{and} \bibinfo{author}{\bibfnamefont{T.~C.}
  \bibnamefont{Killian}}, \bibinfo{journal}{{ Phys. Rev. A}}
  \textbf{\bibinfo{volume}{67}}, \bibinfo{pages}{011401}
  (\bibinfo{year}{2003}).

\bibitem[{\citenamefont{Murillo}(2001)}]{mur01}
\bibinfo{author}{\bibfnamefont{M.~S.} \bibnamefont{Murillo}},
  \bibinfo{journal}{{Phys. Rev. Lett.}} \textbf{\bibinfo{volume}{87}},
  \bibinfo{pages}{115003} (\bibinfo{year}{2001}).

\bibitem[{ele({\natexlab{a}})}]{electrontempgammenote}
\bibinfo{note}{For characterizing dynamics according to initial conditions,
  $\Gamma_e$ is calculated using initial peak density and the naive assumption
  of initial $T_e=2E_e/3k_\mathrm{B}$. The plasma may be far from thermal
  equilibrium at these early times.}

\bibitem[{\citenamefont{Bergeson and Spencer}(2003)}]{bsp03}
\bibinfo{author}{\bibfnamefont{S.~D.} \bibnamefont{Bergeson}} \bibnamefont{and}
  \bibinfo{author}{\bibfnamefont{R.~L.} \bibnamefont{Spencer}},
  \bibinfo{journal}{{ Phys. Rev. E}} \textbf{\bibinfo{volume}{67}},
  \bibinfo{pages}{026414} (\bibinfo{year}{2003}).

\bibitem[{ele({\natexlab{b}})}]{electrontempvirmsnote}
\bibinfo{note}{This is rigorously true for a self-similar expansion and uniform
  ion temperature. It is a good approximations in the absence of these
  conditions \cite{kcg05}.}

\bibitem[{\citenamefont{Dorozhkina and Semenov}(1998)}]{dse98}
\bibinfo{author}{\bibfnamefont{D.~S.} \bibnamefont{Dorozhkina}}
  \bibnamefont{and} \bibinfo{author}{\bibfnamefont{V.~E.}
  \bibnamefont{Semenov}}, \bibinfo{journal}{{ Phys. Rev. Lett.}}
  \textbf{\bibinfo{volume}{81}}, \bibinfo{pages}{2691} (\bibinfo{year}{1998}).

\bibitem[{\citenamefont{Pohl et~al.}(2004{\natexlab{b}})\citenamefont{Pohl,
  Pattard, and Rost}}]{ppr04}
\bibinfo{author}{\bibfnamefont{T.}~\bibnamefont{Pohl}},
  \bibinfo{author}{\bibfnamefont{T.}~\bibnamefont{Pattard}}, \bibnamefont{and}
  \bibinfo{author}{\bibfnamefont{J.~M.} \bibnamefont{Rost}},
  \bibinfo{journal}{{ Phys. Rev. Lett.}} \textbf{\bibinfo{volume}{92}},
  \bibinfo{pages}{155003} (\bibinfo{year}{2004}{\natexlab{b}}).

\bibitem[{\citenamefont{Baitin and Kuzanyan}(1998)}]{bku98}
\bibinfo{author}{\bibfnamefont{A.~V.} \bibnamefont{Baitin}} \bibnamefont{and}
  \bibinfo{author}{\bibfnamefont{K.~M.} \bibnamefont{Kuzanyan}},
  \bibinfo{journal}{J. Plasma Phys.} \textbf{\bibinfo{volume}{59}},
  \bibinfo{pages}{83} (\bibinfo{year}{1998}).

\bibitem[{\citenamefont{Kovalev and Bychenkov}(2003)}]{kby03}
\bibinfo{author}{\bibfnamefont{V.~F.} \bibnamefont{Kovalev}} \bibnamefont{and}
  \bibinfo{author}{\bibfnamefont{V.~Y.} \bibnamefont{Bychenkov}},
  \bibinfo{journal}{Phys. Rev. Lett.} \textbf{\bibinfo{volume}{90}},
  \bibinfo{eid}{185004} (\bibinfo{year}{2003}).

\bibitem[{\citenamefont{Laha et~al.}(2007)\citenamefont{Laha, Gupta, Gao,
  Simien, Castro, and Killian}}]{lgg07}
\bibinfo{author}{\bibfnamefont{S.}~\bibnamefont{Laha}},
  \bibinfo{author}{\bibfnamefont{P.}~\bibnamefont{Gupta}},
  \bibinfo{author}{\bibfnamefont{H.}~\bibnamefont{Gao}},
  \bibinfo{author}{\bibfnamefont{C.~E.} \bibnamefont{Simien}},
  \bibinfo{author}{\bibfnamefont{J.}~\bibnamefont{Castro}}, \bibnamefont{and}
  \bibinfo{author}{\bibfnamefont{T.~C.} \bibnamefont{Killian}},
  \bibinfo{journal}{{submitted}}  (\bibinfo{year}{2007}).

\bibitem[{ele({\natexlab{c}})}]{electrontemptlnote}
\bibinfo{note}{Theshold lowering is sometimes also considered, but electron DIH
  includes this effect because it is related to spatial correlations between
  particles.}

\bibitem[{\citenamefont{Killian et~al.}(2001)\citenamefont{Killian, Lim, Kulin,
  Dumke, Bergeson, and Rolston}}]{klk01}
\bibinfo{author}{\bibfnamefont{T.~C.} \bibnamefont{Killian}},
  \bibinfo{author}{\bibfnamefont{M.~J.} \bibnamefont{Lim}},
  \bibinfo{author}{\bibfnamefont{S.}~\bibnamefont{Kulin}},
  \bibinfo{author}{\bibfnamefont{R.}~\bibnamefont{Dumke}},
  \bibinfo{author}{\bibfnamefont{S.~D.} \bibnamefont{Bergeson}},
  \bibnamefont{and} \bibinfo{author}{\bibfnamefont{S.~L.}
  \bibnamefont{Rolston}}, \bibinfo{journal}{{ Phys. Rev. Lett.}}
  \textbf{\bibinfo{volume}{86}}, \bibinfo{pages}{3759} (\bibinfo{year}{2001}).

\bibitem[{\citenamefont{Pohl et~al.}(2005)\citenamefont{Pohl, Pattard, and
  Rost}}]{ppr05JPB}
\bibinfo{author}{\bibfnamefont{T.}~\bibnamefont{Pohl}},
  \bibinfo{author}{\bibfnamefont{T.}~\bibnamefont{Pattard}}, \bibnamefont{and}
  \bibinfo{author}{\bibfnamefont{J.}~\bibnamefont{Rost}}, \bibinfo{journal}{J.\
  Phys.\ B} \textbf{\bibinfo{volume}{38}}, \bibinfo{pages}{S343}
  (\bibinfo{year}{2005}).

\bibitem[{\citenamefont{Bethe and Salpeter}(1977)}]{bsa77}
\bibinfo{author}{\bibfnamefont{H.~A.} \bibnamefont{Bethe}} \bibnamefont{and}
  \bibinfo{author}{\bibfnamefont{E.~E.} \bibnamefont{Salpeter}},
  \emph{\bibinfo{title}{Quantum Mechanics of One- and Two-Electron Atoms.}}
  (\bibinfo{publisher}{Plenum}, \bibinfo{address}{New York},
  \bibinfo{year}{1977}).

\bibitem[{\citenamefont{Hahn}(1997)}]{hah97}
\bibinfo{author}{\bibfnamefont{Y.}~\bibnamefont{Hahn}}, \bibinfo{journal}{{
  Phys. Lett. A}} \textbf{\bibinfo{volume}{231}}, \bibinfo{pages}{82}
  (\bibinfo{year}{1997}).

\bibitem[{\citenamefont{Hahn}(2000)}]{hah00}
\bibinfo{author}{\bibfnamefont{Y.}~\bibnamefont{Hahn}}, \bibinfo{journal}{{
  Phys. Lett. A}} \textbf{\bibinfo{volume}{264}}, \bibinfo{pages}{465}
  (\bibinfo{year}{2000}).

\bibitem[{\citenamefont{Maiorov et~al.}(1998)\citenamefont{Maiorov, Tkachev,
  and Yakovlenko}}]{mty98}
\bibinfo{author}{\bibfnamefont{S.~A.} \bibnamefont{Maiorov}},
  \bibinfo{author}{\bibfnamefont{A.~N.} \bibnamefont{Tkachev}},
  \bibnamefont{and} \bibinfo{author}{\bibfnamefont{S.~I.}
  \bibnamefont{Yakovlenko}}, \bibinfo{journal}{{Physica Scripta}}
  \textbf{\bibinfo{volume}{51}}, \bibinfo{pages}{498} (\bibinfo{year}{1998}).

\bibitem[{\citenamefont{Killian et~al.}(2005)\citenamefont{Killian, Chen,
  Gupta, Laha, Martinez, Mickelson, Nagel, Saenz, and Simien}}]{kcg05}
\bibinfo{author}{\bibfnamefont{T.~C.} \bibnamefont{Killian}},
  \bibinfo{author}{\bibfnamefont{Y.~C.} \bibnamefont{Chen}},
  \bibinfo{author}{\bibfnamefont{P.}~\bibnamefont{Gupta}},
  \bibinfo{author}{\bibfnamefont{S.}~\bibnamefont{Laha}},
  \bibinfo{author}{\bibfnamefont{Y.~N.} \bibnamefont{Martinez}},
  \bibinfo{author}{\bibfnamefont{P.~G.} \bibnamefont{Mickelson}},
  \bibinfo{author}{\bibfnamefont{S.~B.} \bibnamefont{Nagel}},
  \bibinfo{author}{\bibfnamefont{A.~D.} \bibnamefont{Saenz}}, \bibnamefont{and}
  \bibinfo{author}{\bibfnamefont{C.~E.} \bibnamefont{Simien}},
  \bibinfo{journal}{{J. Phys. B: At. Mol. Opt. Phys.}}
  \textbf{\bibinfo{volume}{38}}, \bibinfo{pages}{351} (\bibinfo{year}{2005}),
  \bibinfo{note}{(Equations 7, 10, 11, and 17 should be multiplied by
  $\gamma_0/\gamma_{eff}$.)}.

\end{thebibliography}

\end{document}